\magnification=1200

\pageno=1
\centerline {\bf ${\cal W}$ GEOMETRY FROM FEDOSOV'S} 
\centerline {\bf  DEFORMATION QUANTIZATION   }
\medskip
\centerline { Carlos Castro }
\centerline { Center for Theoretical Studies of Physical Systems,}
\centerline {Clark Atlanta University, Atlanta, Georgia 30314}
\centerline { }

\smallskip
\centerline { January, 1998} 
\smallskip
\centerline {\bf ABSTRACT}
\bigskip

A geometric derivation of  $W_\infty$ Gravity based on Fedosov's
deformation quantization of symplectic manifolds is presented. 
To lowest order in Planck's constant  it agrees with 
Hull's geometric formulation of classical nonchiral $W_\infty$ Gravity. 
The fundamental object is a ${\cal W}$-valued  connection one form belonging  to the 
exterior algebra of the Weyl algebra bundle associated with the  symplectic manifold.   
The ${\cal W} $-valued  analogs  of the Self Dual Yang Mills equations, obtained from a zero 
curvature condition, naturally lead to the Moyal Plebanski equations, furnishing   
Moyal deformations of self dual gravitational backgrounds associated with the  
complexified cotangent space of a two dimensional Riemann surface. Deformation quantization of $W_\infty$
Gravity is retrieved upon the inclusion of all the $\hbar$ terms appearing in the 
Moyal bracket. Brief comments on Non Commutative Geometry and M(atrix)theory are made.


Subj. Class : Differential Geometry


\bigskip
\bigskip
\bigskip

\centerline { \bf I. Introduction}
\bigskip
Since the classic work of Bayen et al [1] Moyal deformation quantization techniques [2] are starting to become very relevant 
in the area of Non Commutative Geometry, for example  ( see [14] for a current review). 
$W_\infty$ algebras, strings, gravity, etc   were very popular candidates to  extensions of the  ordinary two-dimensional Conformal Field Theory (CFT)  description of strings based on Kac-Moody and Virasoro algebras . For an extensive review on higher conformal spin extensions of CFT we refer to the Physics Reports article [ 5].  Not long ago, the author [8] was able to show that non-critical $W_\infty$ strings are devoid  of BRST anomalies  for target spacetimes of dimension 
$D=27$. The supersymmetric case yielded   $D=11$ and we suggested that the 
an anomaly free (super) membrane should support  non-crititical $W_\infty$ strings in their spectrum.  

$W_\infty$ covariance is of crucial importance in the Weyl-Wigner-Groenewold-Moyal quantization process  [4] and for this purpose its relevance should be investigated further. The role of $W_\infty$ algebras in  Moyal quantization  
was also investigated by [17]. The authors in [24] were able to realize  that nonlinear $W_\infty$ algebras could be obtained by Moyal deformations of the classical $w_\infty$ linear ones. Other important connections among membranes and strings were raised in [25]. 

In this letter we present a very natural framework,  in the language of Fedosov's deformation quantization [3],  where the original formulation of $W_\infty$ Geometry  presented by Hull [6] can be incorporated in a very simple fashion. 
In section {\bf II} we discus Fedosov's Geometry and Moyal SDYM theories in 
$R^4$. In {\bf III} we review Hull's work and show how his formulation admits a straightforward incorporation in the language of Moyal-Fedosov quantization. In the final concluding section we 
outline some of the many applications that deformation quantization techniques have in the theory of extended objects.

\bigskip

\centerline{\bf II. Fedosov Geometry and Moyal Self Dual Yang Mills} 
\smallskip  
\centerline {\bf 2.1.  Fedosov's Deformation Quantization   }
\bigskip

In essence, Fedosov's [3,33]  deformation quantization ; i.e deformations of the Poisson Lie algebra structure on a symplectic manifold, is a generalization of the 
Bayen et al [1] and 
the Weyl, Wigner, Groenewold, Moyal 
(WWGM)[2]  deformation quantization on  flat phase spaces. Given a symplectic manifold, 
${\cal M}$ of dimension $2n$,  with a symplectic globally defined non-degenerate two-form $\omega$, allows  to define a symplectic structure in each tangent space, $T_x {\cal M}$ to the point $x$. The Weyl algebra, ${\cal W}_x$ corresponding to the symplectic space,  $T_x {\cal M}$, is the associative non-commutative algebra over {\bf C} with a unit element. The elements of the Weyl agebra are defined by a formal power series :

$$a(y)=\sum h^k a_{k, i_1......i_l} y^{i_1}.....y^{i_l}.~~~2k+l\ge 0. \eqno (1)$$
where $h$ is a formal parameter which can be identified with $\hbar$ and the coordinates ,
$y^{1},....y^{2n}\in T_x {\cal M}$ are associated to a tangent vector at the point $x$. 
The degrees $1,2$ are prescribed for the variables $y,h$ respectively, so one is summing in  eq-(1)  over
 elements of the algebra of non-negative degree. 
The noncommutative product on the Weyl algebra, $W_x$, which determines its  associative character, 
of two elements, $a(y), b(y)$  is defined :

$$a(y)~o~b(y)=\sum_{k=0}^\infty ({i\hbar \over 2})^k {1\over k!} \omega^{i_1j_1}....\omega^{i_k j_k} 
(\partial_{i_1......i_k} a)(\partial_{j_1......j_k} b). \eqno (2) $$ 
where $\omega^{ij}$ are the components of the inverse tensor of  $\omega_{ij}$, the sympectic two-form.

 Having defined the Weyl algebra at a point $x\in {\cal M}$ one can defined an algebra bundle structure by 
taking disjoint unions of the Weyl algebras at each point; i.e a fibrewise addition. In this fashion a Weyl algebra bundle, ${\cal W} $,  is constructed as $x$ runs over all the points in ${\cal M}$. This allows to build in a set of sections, 
${\cal E}  ({\cal W})$ , denoted by 
$a(x,y,\hbar)$ which  can be written in a power series :

$$a(x,y,\hbar)=\sum h^k a_{k, i_1......i_l}(x)  y^{i_1}.....y^{i_l}.~~~2k+l\ge 0. \eqno (3)$$
where now $a_{k, i_1......i_l}(x)$ are a set of smooth functions defined on the symplectic manifold. 
Differential forms can be constructed whose elements are ${\cal W}$-valued ( instead of the customary forms taking values in a Lie algebra). A ${\cal W}$-valued $q$ differential form may be written :

$$A(x,y,\hbar)=\sum h^k a_{k; i_1......i_l;j_1.....j_q}(x)  y^{i_1}.....y^{i_l}dx^{j_1}\wedge....\wedge dx^{j_q}
.~~~2k+l\ge 0. \eqno (4)$$
and an exterior algebra extension of the Weyl algebra bundle, ${\cal E} ({\cal W}\otimes \Lambda)$ is
 obtained by constructing a deformation  of the wedge product , $\wedge^{o}$. Differential  operators analogous to the exterior derivative and  its dual  ( divergence) are defined in [3] and reviewed in [10,33]. We shall not repeat it here.

A torsion-free symplectic connection, $\partial$,  preserving the symplectic structure can also be defined 
The symplectic connection does $not$ change the degree of the Weyl algebra bundle. 
 $\partial : {\cal E}  (  {\cal W}_p \otimes \Lambda^q) \rightarrow   {\cal E}  (  {\cal W}_p \otimes \Lambda^{q+1})$.  
Fedosov defined a more general connection , $D$, the analog of a gauge covariant derivative, acting in an element, $a$, of the Weyl algebra bundle as :

$$Da=\partial a +{1\over i\hbar} [\gamma, a].~~~\gamma \in {\cal E}  (  {\cal W} \otimes \Lambda^1). \eqno (5)$$
for $\gamma$ a gobally defined ${\cal W} $-valued one form.  Since every symplectic manifold can be equipped with an almost complex structure, $J$, such that the tensor defined by $g(X,Y)=-\omega (JX,Y)$ for all vector fields, $X,Y$, is a Riemannian metric on ${\cal M}$. This yields the curvature 
$R$ associated with the symplectic connection : $R={1\over 4} R_{ijkl}y^iy^j dx^k\wedge dx^l$. The commutator is defined as :
 $[a,b] =a\wedge^ob -(-1)^{qp} b\wedge^oa$ where $a,b$ are respectivley ${\cal W}$ valued 
$q,p$ differential forms.  
The curvature of the connection , $D$, satisfies the propery  :

$$ D^2 a ={1\over i\hbar} [\Omega, a].~~~\Omega =R +\partial \gamma +{1\over i\hbar} \gamma^2. \eqno (6)$$
A connection $D$ is Abelian if, and only if, for any section , $a \in {\cal E}  (  {\cal W} \otimes \Lambda^1)$,~$D^2a=0$. 
Hence from (6) one can infer that the curvature of every  abelian connection, $\Omega$,  is central ( commutes with all the elements $a$ of the Weyl algebra )  and is proportional to
$\omega_{ij}dx^i\wedge dx^j$. Abelian connections are very relevant in the construction of the algebra of Quantum Observables, ${\cal E} ({\cal W}_D) $, 
which is  the subalgebra of the Weyl algebra bundle  comprised of $flat~sections$, 

$$a\in {\cal E} ( {\cal W}_D)\Rightarrow D^2 a=0. \eqno (7)$$
these flat sections ( with respect the Abelian connection) generate a subalgebra of the Weyl algebra bundle called the algebra of Quantum Observables. It is the 
analogous of BRST invariant states in string theory.

Finally one may establish the relationship with the ordinary Moyal star product . 
Fedosov showed that one can assign to a flat section ( relative to an Abelian connection)  
$a (x,y,\hbar)$ a central element of the Weyl algebra , $a_o (x,y=0,\hbar)  \in {\cal Z} $, and viceversa. 
The bijection map ( which we shall not go into details)  and the star product in the central elements of the algebra, $a_o, b_o \in {\cal Z} $ are   :

$$\sigma : {\cal W}_D \rightarrow {\cal Z}.~~\sigma (a) =a_o.~~\sigma^{-1} a_o =a.~~a_o*b_o =\sigma ( \sigma^{-1} (a_o)~ o ~\sigma^{-1} (b_o))=\sigma (a~o~b). \eqno (8)$$
in this way one can construct a  star product in the space of central sections, ${\cal Z} $,   inherited from the 
noncommutative associative Fedosov fibrewise product of  flat sections belonging to the subalgebra of Quantum Observables. 
When the symplectic manifold is flat, $R^{2n}$ , the star product reduces then to the ordinary Moyal star product as expected. In this particular case the tangent bundle is trivial , $T(R^{2n})\sim R^{2n}\times R^{2n}$ and 
there is no difference between the fibrewise ( $y^i$) Fedosov products and the base space $(x^n)$ Moyal ones.  
For definitions of the trace of Quantum Observables in the Weyl algebra bundle, 
its inner automorphisms ( analogous to gauge transformation) , symplectic diffs,... we refer to the literature [3,10].
\bigskip
\centerline {\bf 2.2 Moyal Self Dual Yang Mills} 
\bigskip
It has been known for some time to the experts that $4D$ Self Dual Gravity can be obtained from $SU(\infty)$ SDYM , an effective $6D$ theory, by dimensional reduction. We refer to 
[9] for an extensive list of references. The bosonic and supersymmetric case was studied also by the author [7]. Moyal deformations of 
Self Dual Gravity  were proposed by Strachan [18]  and  rotational Killing symmetry reductions furnished the Moyal quantization of the continuous Toda field [7] . Generalized Moyal Nahm and continuous Moyal Toda equations were developed by [27] based on the work of the SDYM equations by Ivanova and Popov [41]. $W_\infty$ is a natural  symmetry of these integrable theories. Takasaki [19]  and Strachan, among many others,  have emphasized the importance of higher dimensional integrable theories.

In this section we will write down the main equations of Moyal deformations of SDYM 
theories 
in $R^4$ leaving all the technical details for the reference [9,10]. The basic equations are obtained from a zero curvature condition which allows to gauge two of the fields $A_x, A_y =0$ and yields for the remaining third equation   :

$$\partial_{\tilde x} A_{\tilde y} -\partial_{\tilde y } A_{\tilde x} + 
[A_{\tilde y}, A_{\tilde x} ] =0. \eqno (9)$$
A WWGM quantization of a $SU(N)$ SDYM requires finding a representation of the Lie algebra $su(N)$ in a suitable Hilbert space, where the ${\hat A}_\mu$ are vector-valued operators living in a Hilbert space. A WWGM quantization requires the construction of  operators acting in the Hilbert space of square integrable 
functions on the line, $L^2 (R)$, and the use of the symbol WWGM map to define a one-to-one correspondence  of self adjoint operators into real valued smooth functions in  the phase space $q,p$ associated with the line. Evenfurther, the WWGM map takes  commutators, ${1\over i\hbar}[{\hat A}, {\hat B}]$  into Moyal Brackets. In [27] we have shown that in general one should  enlarge the phase space with the introduction of $q^i,p^i$  to accomodate for the Moyal deformations of continuum Lie algebras introduced by Saveliev and Vershik [42] .
Hence eq-(9) becomes an effective  $6D$ equation after the WWGM quantization   :

$$\partial_{\tilde x} A_{\tilde y} -\partial_{\tilde y } A_{\tilde x} + 
\{A_{\tilde y}, A_{\tilde x} \}_{Moyal} =0. \eqno (10)$$
where now, $A_\mu (x,y,{\tilde x}, {\tilde y},q,p,  \hbar)$. The last equation admits,  in the $\hbar \rightarrow 0$ limit,  reductions 
to many of the known integrable equations, like the Plebanski first and second heavenly equations, the Park-Hussain and  Grant equations,....[9 ] . Furhermore, eq-(10) may be obtained directly from a Lagrangian [9]. In fact, dynamics of higher spin fields can be encoded in zero curvature constraints. This has been explained in particular by Vasiliev [40]. The $6D$ Moyal SDYM 
equations (10) admit a reduction to $4D$ as follows :

$$A_{\tilde y}=\partial_x \Theta =  \partial_x \Theta' -{1\over 2} {\tilde x} . ~~~
A_{\tilde x}=-\partial_y \Theta = - \partial_y \Theta' +{1\over 2} {\tilde y}. \eqno (11)$$
with the $6D$ function $\Theta$ :

$$\Theta (x,y,{\tilde x}, {\tilde y},q,p \hbar)
\equiv \Theta'(x+{\tilde y}, {\tilde x} -y, q,p,\hbar) -{1\over 2} (x{\tilde x} +y {\tilde y} ). \eqno (12)$$
Inserting (11,12) into (10) yields the $4D$ Moyal first heavenly Plebanski equation  :

$$\{\Omega_{w}, \Omega_{\tilde w} \}_{Moyal} =1.~~~\Omega (w, {\tilde w},q,p,\hbar)\equiv \Theta'.  ~~~w=x+{\tilde y} ;~ {\tilde w} ={\tilde x} -y                     \eqno (13) $$ 
Eq-(13) can also be rewritten, due to the fact that the Moyal bracket of the variables 
$x,y, {\tilde x},{\tilde y}$ is zero, :

$$\{ \partial_x \Theta ( x,y,{\tilde x}, {\tilde y},q,p , \hbar), \partial_y \Theta ( x,y,{\tilde x}, {\tilde y},q,p,  \hbar)
\}=1. \eqno (14)$$

The dimensional reduction of the Moyal SDYM theory, from $6D\rightarrow 4D$,  leading to 
the Moyal Plebanski equation , can be interpreted as a foliation of the $6D$ space into 
$4D$ leaves endowed with Moyal deformed ( self dual/anti self dual curvatures) hyper-Kahler Ricci flat  metrics and  parametrized by the coordinates ${\tilde x}, {\tilde y} $ of the $6D$ space  :

$$\{  \partial_x K_{\tilde x, \tilde y} (x,y,q,p,\hbar),\partial_y  K_{\tilde x \tilde y} (x,y,q,p,\hbar) \}_{Moyal} =1. \eqno (15)$$
where $ K_{\tilde x \tilde y} (x,y,q,p,\hbar)$ is a two-parameter family of Moyal deformed Kahler potentials. This is attained by setting in (14,15) :

$$\partial_y  K_{\tilde x \tilde y} (x,y,q,p,\hbar)= 
\partial_y \Theta ( {\tilde x}, {\tilde y}|x,y, q,p,  \hbar). ~
\partial_x  K_{\tilde x \tilde y} (x,y,q,p,\hbar)= 
\partial_x \Theta ( {\tilde x}, {\tilde y}|x,y, q,p, \hbar).\eqno (16)$$
Hence, for running values of  ${\tilde x}={\tilde x}_o; {\tilde y}={\tilde y}_o$ which characterize  the foliation, eqs-(13,14,15) yield  a two-parameter family of 
Moyal heavenly metrics  encoded in the Moyal Plebanski potential :  $\Omega (w, {\tilde w},q,p,\hbar)$. The foliation is represented by the two-parameter family of four-dimensional noncommutative manifolds, $X^4_{{\tilde x, \tilde y}}(x,y,q,p,\hbar)$, since the star product of two functions in phase space is noncommutative. The connections between Non 
Commutative Geometry, Matrix models [11,12,14,15]  and String theory [13]  is now being developed by a large number of authors. We apologize for excluding many relevant references. The relevance of Self Dual Gravity in connection with $N=2$ strings was initiated by [20] and pursued by many others [30,31]. Important remarks about M (atrix) models and $N=2$ strings have appeared in [29] .  The role of self dual gravity and $W_\infty$ algebras was initially 
worked out, among others, by [21,22,23]. This completes this short review about Moyal SDYM in $R^4$.     
\bigskip
\centerline {\bf III. {\cal W}-Geometry From Fedosov } 
\centerline {\bf 3.1  Hull's Formulation  of  $W_\infty$ Gravity} 
\bigskip
Sometime ago , Hull [6] with great insight, presented a geometric formulation of $W_\infty$ Geometry as a gauge theory of the group of symplectic diffeomorphisms of the cotangent bundle of a two-dimensional Riemann surface, $Diff_o (T^*{\cal N})$. The infinite set of symmetric gauge tensor fields are
 ${\tilde h}_{(n)} ^{\mu_1....\mu_n}$, $n=2,3,...\infty$ ;  $\mu=0,1$. These transform as densities   under ${\cal W}$ gauge tranformations  :

$$\delta {\tilde h}_{(s)}^{\mu_1....\mu_s} =\sum_{m,n} \delta_{m+n,s+2} [(m-1)
\lambda^{ (\mu_1\mu_2...}_{(m)}\partial_\nu {\tilde h}_{(n)} ^{...\mu_s)\nu} 
-(n-1) {\tilde h} ^{ \nu(\mu_1\mu_2...}_{(n)}\partial_\nu {\lambda}_{(m)} ^{...\mu_s)} $$
$$+ {(m-1) (n-1) \over p-1} \partial_\nu \{
\lambda^{ \nu (\mu_1\mu_2...}_{(m)} {\tilde h}_{(n)} ^{...\mu_s)}
 -{\tilde h}^{ \nu (\mu_1\mu_2...}_{(n)} {\lambda}_{(m)} ^{...\mu_s)}  \}]. \eqno(17) $$ 
$s$ labels the conformal spin $2,3,4....\infty$ of the gauge fields whose physical components have helicity $\pm s$. From 
the  generalized  notion of a scalar line element 
$$ds =(g_{\mu_1.....\mu_n} dx^{\mu_1} ....dx^{\mu_n} )^{(1/n)}$$  
Hull proposed an action of the form  :

$$ S=\int d^2x {\tilde  F} (x,y). ~~~  
{\tilde  F} (x,y)=\sum_{n=2}^\infty {1\over n} {\tilde h}_{(n)}^{\mu_1....\mu_n} y_{\mu_1}.....y_{\mu_n}. \eqno (18)$$
where the function ${\tilde F} (x,y)$ is a co-metric ${\cal W}$-density in $d=2$ instead of a ${\cal W}$-scalar. The action represents the integrated  generalized world interval  along a  section of the bundle $T^* {\cal N}$,  where the fiber coordinates , $y_\mu$, when restricted to a  section, $\Sigma$,  can be interpreted as the gradients of the matter fields $y_\mu|_\Sigma =y_\mu (x)=\partial_\mu \phi (x)$. 
Hull used one and only one bosonic scalar field, $\phi (x)$ , living in the two-dimensional world sheet. Later we will see how to include a set of matter fields, $\phi^i (x)$ representing the embedding spacetime coordinates of the string worldsheet.   
The nonlinear transformation property of the field $\phi (x^\mu)$ is :

$$\delta \phi =\Lambda (x^\mu, y_\mu) =\sum_{n=2}^\infty {\lambda}_{(n)}^{\mu_1....\mu_n}(x^\mu)  y_{\mu_1}.....y_{\mu_n}. \eqno (19)$$     
However, the formulation based on the infinite number of fields ${\tilde h}_{(s)} ^{\mu_1...\mu_s}$ was redundant ( the action was reducible) in the sense that there are more gauge fields than are needed. For example,  ${\tilde h}^{\mu\nu}$ has three independent components for only two gauge symmetries. Hull proved that a gauge invariant 
( invariant under the transformations given by (17,19) )  constraint could be consistently imposed on all the gauge fields in such a way that one could recast the action solely in terms of a set of unconstrained fields, $h_{(s)}^{\mu_1....\mu_s}$ with their traces removed at the linearized level 
 after exploiting the ${\cal W}$-Weyl conformal invariance. 
Thus, at the $linearised$ level  with respect to a flat $\eta_{\mu\nu}$ two-dim metric  and to lowest order in the gauge fields one has :

$${\tilde h} ^{\mu_1...\mu_s}_{(s)} = [{h} ^{\mu_1...\mu_s}_{(s)}-traces ]+O(h^2)...$$
$$\delta {\tilde h} ^{\mu_1...\mu_s}_{(s)}= \partial ^{(\mu_1} {\lambda} ^{\mu_2...\mu_s )}_{(s)}+.....; ~~~{\lambda} ^{\mu_1...\mu_{s-1} }_{(s)}= [{k} ^{\mu_1...\mu_{s-1} }_{(s)} -traces]+O(k^2).... \eqno (20)$$  
The gauge-invariant background independent master constraint that generates $all$  the constraints on the gauge fields, upon expansion in powers of $y_\mu$,  and which allows to recast  the action solely in terms of the unconstrained fields  is  : 

$$det [{\tilde G} ^{\mu\nu} (x^\mu, y_\mu) ]=1.~~~{\tilde G} ^{\mu\nu} (x^\mu, y_\mu)=
{\partial^2 {\tilde F} (x^\mu, y_\mu) \over \partial y_\mu \partial y_\nu} . \eqno (21)$$
For $2+2$ signature one has $-1$ for the determinant instead in the r.h.s of (21). 
The constraint (21)  can be solved in a particular way by recurring to the ${\cal W}$-Weyl conformal invariance which allows to gauge away all the traces  of the unconstrained fields, leaving only traceless fields in the action with helicities $\pm s$.

The geometrical significance of the constraint (21) is the following . If one sets $y_\mu =z_\mu +{\bar z}_\mu$ where $z_\mu$ are the complex coordinates on $R^4$, $\mu=0,1$,  
allows to view $(x^\mu, z_\mu, {\bar z}_\mu)$ as the coordinates for the bundle   
{\bf C} $T^*_x {\cal N}$ , the complexification of the cotangent space $T^*_x {\cal N}$ at a  point $x \in {\cal N} $. The co-metric  function is then reinterpreted as     :

$${\tilde F} (x^\mu, y_\mu) =K_x (z, {\bar z})={\tilde F}(x^\mu , z_\mu +{\bar z}_\mu ). \eqno (22) $$ 
with $K_x (z, {\bar z})$ the Kahler potential depending on the combination $z_\mu +{\bar z}_\mu $ which is tantamount of a Killing symmetry reduction condition, the metric does not depend on the two imaginary components of $z_\mu$.  :

$${\partial^2 K_x (z, {\bar z}) \over \partial z_\mu \partial {\bar z} _\nu }=   
 {\partial^2 {\tilde F} (x^\mu, y_\mu) \over \partial y_\mu \partial y_\nu} .\eqno (23) $$
The gauge invariant ( under the transformations given by (17,19) ) master constraint (21) was  equivalent to the Monge-Ampere equation ( Plebanski equation) for a Ricci flat, hyper-Kahler manifold associated with the compexified cotangent space at each point $x^\mu$, of the original two-dimensional surface, world-sheet, ${\cal N}$ :  {\bf C}$T^*_x {\cal N} \sim $ {\bf C}$^2$.  
This implies that for each $x^\mu$, the corresponding curvature tensor is either self-dual or antiselfdual. Summarizing, for each, $x^\mu \in {\cal N} $, 
${\tilde F} (x^\mu, y_\mu =z_\mu +{\bar z}_\mu ) =K_x (z, {\bar z})$ 
is the Kahler potential for a hyperKahler metric on $R^4$ with two commuting tri-holomorphic 
Killing vectors.  
Hence,  $K_x (z, {\bar z})$ furnishes a two-parameter family of metrics labelled by the points 
$x^\mu \in {\cal N}$ and a bundle over ${\cal N}$ is obtained whose fibers at each point  are 
isomorphic to {\bf C}$^2$ and  equipped with a half-flat metric with two Killing vectors. 

When $d=2$ it is  possible to construct invariant actions under a subgroup of the ${\cal W} $ transformations ( symplectic )  if, and only if, a ${\cal W} $ scalar-density exists, ${\tilde F} (x^\mu, y_\mu)$ so that the
action is ${\cal W}$-invariant up to surface terms . In $d>2$ no such density exists.  
The reason that the action was solely invariant  under a subgroup of   
$Diff_o (T^*{\cal N})$  was due to the fact that a constraint on the gauge parameters $\lambda _{(s)}^{\mu_1....\mu_s}$ had to be imposed as well 
because under ${\cal W}$ transformations given by eqs-(17,19),  the action behaves like :

$$\delta S =\int d^2x \partial_\mu (\Omega^\mu +X) =0 \Rightarrow X=0. \eqno (24)$$
To $lowest$ order in $\Lambda (x^\mu, y_\mu)$ the constraint $X=0$ reads :

$$det[{\partial^2 ({\tilde F}+\Lambda) (x^\mu, y_\mu) \over \partial y_\mu \partial y_\nu}]=1. \eqno (25)$$
for $2+2$ signature one has $-1$ on the r.h.s of (25). The above constraint represents $infinitesimal$  deformations of the hyper-Kahler geometry with two-Killing vectors, for  a deformed Kahler potential ${\tilde F} \rightarrow {\tilde F} +\Lambda$ .  
The constraint $X=0$ on the gauge parameters is not fully symplectic-diffs invariant like eq- (21) was, it is only invariant  under a subgroup of the symplectic-diffs [6] . In the next section we will present a straighforward 
derivation of all these constraints based on Moyal deformations of SDYM.

Summarizing, invariant actions under a subgroup of the symplectic-diffs can be constructed in terms unconstrained gauge fields  ,  $h_{(s)}^{\mu_1.....\mu_s}$ , and transformation laws with gauge parameters, 
$  k_{(s)}^{\mu_1.....\mu_s}$ . Their traces can be removed by means of exploiting the  ${\cal W}$-Weyl conformal invariance. These traceless ( irreducible) fields and parameters appear  in the invariant action and transformation 
laws   $nonlinearly$ in the form of  
${\tilde h} _{(s)}^{\mu_1.....\mu_s} $, and $ \lambda_{(s)}^{\mu_1.....\mu_s}$. The symmetry algebra of nonchiral $W_\infty$ Gravity is a subalgebra of the $Diff_o(T^*{\cal N}) $.                 ,

For references on the twistor transform, reductions to $W_N$ Geometry and its connection to Strominger's special geometry [28], finite versus infinitesimal transformations, etc... we refer to Hull's original work [6]. 
Now we are ready to recast the constraints (21,25) in the language of Fedosov-Moyal quantization. 

\bigskip 
\centerline {\bf 3.2 {\cal W}-Geometry from Moyal-Fedosov } 
\smallskip

From the last two sections we can see that Hull's construction of $W_\infty$ Geometry fits very naturally with the Moyal Self Dual Gravitational equations (14,15) ( after a Killing symmetry reduction) . The $\hbar =0$ limit in  eqs-(14,15) of  the Moyal brackets turns  into Poisson brackets and the latter, in turn,  can be formulated as a simple determinant :

$$\{\partial_x K_{\tilde x, \tilde y}(x,y,q,p,\hbar=0),\partial_y K_{\tilde x, \tilde y}(x,y,q,p,\hbar=0)\}_{Poisson} =1 \Rightarrow $$
$$ det [{\partial^2 K_{\tilde x,\tilde y}(x,y,q,p,\hbar=0)\over \partial \xi^i \partial 
{\bar \xi} ^j}] =1. \eqno (26)$$
where $\xi^1;\xi^2 $ plus complex conjugates conjugates   ${\bar \xi}^1;
{\bar \xi}^2 $ are suitable functions of the $x,y,q,p$ variables. Two Killing symmetry reductions  must be subsequently   performed in such a way that the $x,y,q,p$ dependence appears solely in the combinations  $y^i=\xi^i +{\bar \xi}^i, i=1,2$. In this fashion one can make contact with the two-Killing symmetry reductions imposed by Hull [6].     
The four variables $x,y,q,p$ admit a natural interpretation in terms of the $z_\mu, {\bar z}_\mu$ variables which described the {\bf C}$^2$  fibers of the complexified cotangent space of the two-dim surface, ${\cal N}$  at a given point $x^\mu \in {\cal N}$. The $\hbar =0$ limit of the Moyal heavenly equations associated with the two-parameter family of leaves foliating the $6D$ space, after a further Killing symmetry reduction $y^i=
\xi^i +{\bar \xi}^i$,   admit a direct correspondence with the constraints (21,23) present in Hull's formulation :

$$   K_{\tilde x, \tilde y}(x,y,q,p,\hbar=0)=
K_{\tilde x, \tilde y}(y^i=\xi^i +{\bar \xi}^i,\hbar=0)\leftrightarrow 
K_{x^\mu} (z_\mu, {\bar z}_\mu)=$$
$$K_{x^\mu} (y_\mu =z_\mu +{\bar z}_\mu )={\tilde F} (x^\mu, y_\mu). \eqno (27)$$
Since we were able to identify $ K_{\tilde x, \tilde y}(x,y,q,p,\hbar)$ with a two-parameter family of Moyal Kahler potentials , $ 
\Theta (\tilde x,\tilde y|x,y,q,p,\hbar )$,  it follows  that Hull's co-metric density function can naturally be embedded into $\Theta$ by performing the $double$ infinite summation typical of Fedosov's Geometry and recurring to the two Killing symmetry reductions     :

$$\Theta (\tilde x,\tilde y|x,y,q,p,\hbar) =\sum_{2n+l\ge 0}\hbar^n 
\Theta_{n; i_1,...i_l}~(\xi+  {\bar \xi})^{i_1}....... (\xi+  {\bar \xi})^{i_l}. 
\eqno (28)$$
the coefficients belong to  a one parameter family of smooth tensorial functions  of the ${\tilde x}, {\tilde y} $ variables ( parametrized by the integer $n$)  :  
$$\Theta_{n; i_1,...i_l} ({\tilde x}, {\tilde y}). \eqno (29)  $$
Hull's co-metric density corresponds solely  to the $\hbar^0$ terms :

$$ {\cal F} (x^\mu, y_\mu) \leftrightarrow 
\sum_{l\ge0} (\xi+{\bar \xi})^{i_1}....... (\xi+  {\bar \xi})^{i_l}
\Theta_{0; i_1,...i_l} ({\tilde x}, {\tilde y}).. \eqno (30)$$ 
while  the higher order $\hbar$ corrections implement in one scoop  the Moyal deformation quantization of nonchiral $W_\infty$ gravity. The series in (30) must start with $l=2$ in order to match the expression for the co-metric density. A truncation to zero of the first two terms in the series is necessary.  Futhermore, we have an expansion in  contravariant vectors versus Hull's expansion in terms of covariant vectors. Indices are raised and lowered using the two-dim metric. 
Eq-(30) may be reinterpreted  as the zeroth-order terms ( in $\hbar$) associated with a Fedosov deformation quantization of the symplectic two-dim manifold with coordinates ${\tilde x},{\tilde y}$. The tangent vector at the each point should have for coordinates $y^i=\xi^i+{\bar \xi}^i$. This is relevant to the chiral model approach to self dual gravity [10] where instead of starting with a direct Yang-Mills formulation one writes down the Fedosov deformations of WZNW actions.      

If the afore-mentioned interpretation is adopted, Hull's action will be the integration of the $\hbar^0$ componet of $\Theta$ along a section of the cotangent bundle of the two-dim manifold. Notice that this is $not$ the same as taking the Fedosov trace of the zero form, $\Theta$ , belonging to the exterior algebra of the Weyl algebra bundle associated to the two-dim symplectic manifold ( Riemann surface). Secondly, the $W_\infty$ transformations of the higher spin fields and the matter fields in eqs-(17,19) must not be confused with the Fedosov's noncommutative fibrewise product algebra of the $y^i$ coordinates. Although the $W_\infty$ algebra
can be identified with Moyal deformations of the classical area-preserving diffs [24]. 

Instead of the foliation  picture presented above of the $6D$ space into a family of Ricci flat $4D$ leaves, there is yet another interpretation of the relation of the Moyal SDYM equations and Hull's formulation of $W_\infty$ gravity; i.e the constraints (21) on the Kahler potential. This requires now a dimensional reduction of the original four dimensional spacetime and to extend the construction of {\bf 2.2} to the case that the phase space manifold 
parametrized by $q,p$ is no longer flat. This is attained as follows : 

The gauge fields living on the four-dim manifold, $R^4$, will now take values in the Weyl algebra bundle, ${\cal W}$,  constructed over the curved phase space, $\Gamma$, whose coordinates are $q,p$. Now one has 
${\tilde A}_\mu (X^\mu|q, p,y^1,y^2;\hbar)$ . The zeroth-element is precisely the central section of the algebra ${\cal E}({\cal W}_D)$ obtained from the projection : 
$\sigma ({\tilde A}_\mu)={\tilde A}_\mu (X^\mu|q,p,0,0;\hbar)=A_\mu (X^\mu|q,p;\hbar)$. Fedosov gave the relation which allows one to reconstruct the full ${\tilde A}_\mu$ field from the knowledge of the central section : $A_\mu$. It resembles an expansion in terms of Riemann normal coordinates, see [3,10, 33] for further details. As explained in {\bf 2.1} the bijection map, $\sigma$ yields Moyal star products in the space of  central sections inherited from the Fedosov fibrewise products  of the flat sections.  

Eq-(10) can be obtained from a Lagrangian both in the Moyal and Fedosov case. They represent the deformations of the six dimensional version of the second heavenly equation . The only difference in the Fedosov case is that now one is required to use the definition of the Fedosov trace and to take Fedosov's fibre-wise products instead of Moyal ones. In the flat phase space limit, $\Gamma\rightarrow R^2$, one recovers naturally the Lagrangian of [9] . The trace becomes then an integration w.r.t the $q,p$ coordinates. The original four-dim action will then be an effective $6D$ one  : $M^4\times R^2$ as a result of taking the Fedosov trace :

$$S=\int d^4X \int dqdp ~ [-{1\over 3}\Theta_o *\{ \partial_x \Theta_o,  
\partial_y \Theta_o \} +{1\over 2}(\partial_x \Theta_o)*( \partial_{{\tilde x}}\Theta_o)+
{1\over 2}(\partial_y \Theta_o)*( \partial_{{\tilde y}}\Theta_o)]. \eqno (31) $$ 
with $x,y,{\tilde x}, {\tilde y}$ being the coordinates of $M^4$. The Moyal star product is w.r.t the $q,p$ coordinates of the flat phase space, $R^2$. $\Theta_o (X^4|q,p,\hbar) $ is a scalar field living on $M^4$ and taking values in the space of central sections of the ${\cal E}({\cal W}_D)$ algebra. The action yields the equations of motion for the Moyal deformations of the 
$6D$ version of the second heavenly equation. In the classical $\hbar =0$ limit, Moyal brackets turn into Poisson ones. A dimensional reduction from $6D$ to $4D$ of the type : $\partial_x =\partial_{{\tilde x}}$ and 
$\partial_y =\partial_{{\tilde y}}$ furnishes the Park-Hussain 
second heavenly equations associated  with the chiral model approach to Self Dual Gravity. 
Once again, we see that SDG is an essential geometrical ingredient in these constructions.

Therefore, after an integration of eq-(31) w.r.t the $q,p$ coordinates is performed and a $futher$ dimensional reduction from $M^4$ to two-dimensions is taken one can make contact with 
Hull's action (18). Both views should in principle be equivalent since both lead to 
Self Dual Gravity. One is expressed in terms of the first heavenly form and the other in terms of the second heavenly one. A Darboux transformation relates the first heavenly equation with the second heavenly one ( the chiral models) . Further analysis of these two views : foliations versus reductions and the construction of induced  $W_\infty$ gravity from WZNW models [21,43] and its Fedosov deformation quantization will appear elsewhere.

The other constraint (25) that broke the full symplectic-diffs invariance down to a subgroup can also be understood within the framework of Moyal SDYM. The analogy of gauge transformations of the Moyal YM potentials reads :

$$A_M \rightarrow A_M +  \partial_M \lambda+ \{A_M ,\lambda\}_{Moyal}. ~~~
F_{MN}  \rightarrow    \{F_{MN} , \lambda\}_{Moyal}. \eqno (32a)$$
under gauge transformations the zero curvature ( SDYM) conditions are preserved and hence the Moyal Plebanski equations are naturally gauge invariant exactly like it happened to the constraint (21). 
However, one can see that $\Theta \rightarrow \Theta +\Lambda$ is not always a symmetry of the Moyal SDYM theory. Under shifts in $\Lambda$ the Moyal YM potentials do $not$ transform as (31) but instead :

$$  A_{\tilde x}=- \partial_y \Theta \Rightarrow A_{\tilde x}  
\rightarrow A_{\tilde x}  -\partial_y \Lambda . \eqno (32b)$$
eq-(32) does not represent a true gauge transformation of the 
$A_{\tilde x}$  Moyal YM  potential due to the gauge noncovariant 
$-\partial_y \Lambda$ piece. One would  only have true gauge covariance if, and only if, $\Lambda,\lambda$ satisfy  the property : 
$$ -\partial_y \Lambda = -\partial_y \Lambda -  \{ A_y, \Lambda\}_{Moyal} =
  \partial_{\tilde x}\lambda +\{  A_{\tilde x} , \lambda\}_{Moyal}= \delta A_{\tilde x}\eqno (33) $$
which implies the two conditions  :

$$\{ A_y, \Lambda\}_{Moyal}=0.~~~-D_y \Lambda =D_{\tilde x} \lambda. \eqno (34)$$
this is a very restricted condition on the gauge parameter $\lambda$ and the shift parameter $\Lambda$. It is not surprising that the symplectic-diffs invariance is not fully preserved under shifts in $\Lambda$ of the Moyal Kahler potential $\Theta$. We recall from [9] that $A_y$ could be gauged to zero due to the zero curvature condition that furnishes the Moyal SDYM equations in $R^4$. Hence, the gauge $A_y =0$ obeys one of the conditions. However, there still remains the other condition restricting the parameters $\lambda, \Lambda$ in a highly nonlinear manner  in terms of the remaining Moyal YM potentials. Similar considerations apply for the other $A_{\tilde y}$ Moyal YM potential.

\centerline {\bf IV. Conclusion} 
\bigskip 
We have shown that Hull's formulation  of nonchiral $W_\infty$ Geometry fits in very naturally inside a larger picture : Moyal-Fedosov Geometry. Hull's constraints (21,25) have a natural interpretation in the Fedosov-Moyal quantization program. Furthermore, the inclusion of all the powers in $\hbar$ implements in a straightforward fashion the deformation quantization of nonchiral $W_\infty$ gravity. In general one must consider a Yang-Mills like formulation of ${\cal W}$-geometry based on a ${\cal W}$-valued connection one-form belonging to the exterior algebra of the Weyl-algebra bundle associated with the symplectic manifold.  
Some of the advantages of this formulation of ${\cal W}$-geometry are the following , 

$\bullet$ The incorporation of many bosonic fields $\phi^i, i=1,2,3,.....D$ is straightforward in Fedosov's Geometry : the coefficients in the expansion (1) are  matrix valued; i.e. the coefficients take values in the bundle Hom ($E,E$) where $E$ is a vector bundle over ${\cal M}$. For more details on this see [3]. The scalars $\phi^i$ represent the embedding coordinates  of the world sheet in a target spacetime background of dimension $D$.  Since these fields are reinterpreted as matrix valued sections with a noncommutative fibrewise Fedosov product it is clear why the embedding coordinates of the string world sheet inherit a noncommutative product structure !  

$\bullet$ Quantization of $p$-branes [37] should be achieved using the Zariski product [34] where deformation quantization methods for higher dimensional 
generalizations  of symplectic geometry are achieved : the so-called Nambu-Poisson Hamiltonian Mechanics. Deformation quantization of Poisson manifolds has also  been discussed by [35,36].  

$\bullet$ We hope that the  Moyal-Fedosov deformation quantization approach to ${\cal W}$ Geometry will provide many new insights into the nonperturbative structure of string theory. In particular the role of $W_\infty$ strings [8].  
Paraphrasing Fairlie [16] : Moyal stands for ``M''; to-day  we advocate that 
M-theory ``stands  upside down''   for ${\cal W}$. Important objections  why $W$ geometry is still far from being understood have been recently raised by [39].  

$\bullet$ Zucchini has shown that $4D$ Conformal Field Theory can 
naturally be formulated in real four-folds ( Kulkarni four-folds )  endowed with an integrable quaternionic structure [ 26] and a $4D$ extension of $2D$ CFT on Riemann surfaces was constructed. Quaternionic ( Fueter) analyticity played the role of  $2D$ holomorphic analyticity. It is warranted to explore further the connections between the self duality and  ${\cal W}$-Weyl conformal invariance properties of $W_\infty$ Geometry and Zucchini's Quaternionic Geometry of $4D$ Conformal Field Theory. 
$4D$ generalizations of the $2D$ WZNW models were studied by [32]. Earlier work in that direction was provided by Park [22]. 
The fact that quaternions are noncommutative points in the right direction. The non-associative character of octonions suggests that $8D$ Non Associative Geometries are no longer 
speculative figments of the imagination but should also come to play an important role in physics.      

$\bullet$ The view advocated here of ${\cal W}$ geometry as flat foliations in higher dimensions may have an important relation with Zois [38] proposal for a nonperturbative Lagrangian of $M$  theory in $11D$ in terms of characteristic classes of flat foliations ( although in in odd dimensions) .   

$\bullet$ A forthcoming project involves to write down Yang-Mills types of action characterizing the higher spin field dynamics. In particular to establish the connection to Vasiliev's work [40] : 
Higher-spin gauge theories in four, three and two-dimensions and interactions of matter fields based on deformed oscillator algebras have been studied by Vasiliev and others . A reformulation of the dynamical equations of motion, called ``the unfolded formulation ``,  in a form of a zero curvature condition and a covariant-constancy condition imposed on an infinite collection of zero-forms allowed Vasiliev to describe all spacetime derivatives of all the dynamical fields and reconstruct these fields by analyticity in some neighbourhood of a fixed point. There are many similarities with the work of [40] and ours : a zero curvature condition is imposed; a star product also appears in order to describe the nonlinear dynamics; a deformed oscillator algebra realizing the universal enveloping algebra of symplectic groups is essential .....;
What is required then is to integrate Vasiliev's formulation in the Moyal-Fedosov program.

\bigskip
\centerline {Acknowledgments } 
We are indebted to George Chapline for many discussions .  
\bigskip
\centerline { \bf REFERENCES}

1. F. Bayen, M. Flato, C. Fronsdal, A. Lichnerowicz and D. Sternheimer , 

`` Deformation Theory and Quantization `` Ann. Phys. {\bf 111} (1978) 61. 

2. J. Moyal,  ``Quantum Mechanics as a Statistical Theory''.  

Proc. Cam. Phil. Soc. {\bf 45} (1945) 99

E. Wigner, `` Quantum corrections for thermodynamic equilibrium `` 

Phys. Rev. {\bf 40} (1932) 749. 

H. Groenewold, `` On the principles of elementary quantum mechanics 

`` Physica {\bf 12} (1946) 405.  

H. Weyl, `` Quantum mechanics and the theory of groups'' Z. Phys. {\bf 46} (1927) 1. 

3. B. Fedosov, `` A simple geometrical construction of deformation quantization''   

J. Diff. Geometry, {\bf 40} (1994) 213.  

4. T. Deroli, A. Verciu,   `` $W_\infty$-covariance of the 

Weyl-Wigner-Groenewold-Moyal quantization `` J. Math. Phys. {\bf 38} (11) (1997) 5515. 

5. P. Bouwknegt, K. Schouetens,   `` W-symmetry in Conformal Field Theory''. 

Phys. Reports {\bf 223} (1993) 183-276. hep-th/9210010.

6. C. Hull, `` The Geometry of W-gravity'' Phys. Lett {\bf B 269} (1991) 257. 

``${\cal W}$ -Geometry `` , hep-th/9211113. 

``Geometry and ${\cal W}$-gravity `` hep-th/9301074. 

7. C. Castro,  `` A Moyal quantization of the continous Toda field'' 

Phys. Lett {\bf B 413} (1997) 53. 

`` $SU(\infty)$ (super) gauge theories and self dual ( super) gravity'' .

 J. Math. Phys {\bf 34} (1993) 681. 

`` On $W$ gravity, $N=2$ strings and $2+2$ $SU(\infty)$ Yang-Mills instantons'' . 

J. Math. Phys {\bf 35} (1994) 3013

8. C. Castro, `` $D=11$ supermembrane instantons, $W_\infty$ strings and the 

super Toda molecule''. J. of Chaos, Solitons and Fractals, {\bf 7} (7) (1996) 711, 

`` The non-critical $W_\infty$ string sector of the self dual membrane'' hep-th/9612160. 

9. J.  Plebanski, M. Przanowski, `` The Lagrangian of Self 

Dual Gravity as the limit of the SDYM Lagrangian `` . Phys. Lett {\bf A 212} ( 1996) 22

10.H. Garcia-Compean,  J.  Plebanski and M. Przanowski : `` Geometry associated 

with SDYM and the chiral model approach to Self Dual Gravity''. hep-th/9702046. 

11. A. Connes, M. Douglas, A. Schwarz, ``Non Commutative Geometry and Matrix 

Theory : Compactification on Tori'' hep-th/9711162.

12. P. Ming Ho, Y.S. Wu, `` Non Commutative Gauge Theories in Matrix Models `` . 

hep-th/9801147.

13. F. Lizzi, R. Szabo : ``  Non Commutative Geometry and Spacetime Gauge 

Symmetries of String Theory ``  hep-th/9712206.

14. J. Froehlich, O. Grandjean, A. Recknagel : `` Supersymmetric Quantum Theory, 

Non Commutative Geometry and Gravitation ''. hep-th/9706132.

15.T. Banks, W. Fischler, S. Shenker,  L. Susskind, `` M theory as a Matrix 

theory : a conjecture ``  Phys. Rev. {\bf D 55} (1997) 112.

16. D. Fairlie, `` Moyal Brackets in M Theory''. hep-th/9707190. 

17. E. Gozzi, M. Reuter , `` Quantum Deformed Geometry in Phase Space'' 

DESY-92-191 .

18. I. Strachan, `` The Moyal algebra and integrable 

deformations of the self dual Einstein equations `` . Phys. Lett. {\bf B 283} (1992) 63.

19. K. Takasaki, `` Dressing operator to Moyal algebraic formulation of 

Self Dual gravity'' J. Geom. and  Phys. {\bf 14} (1994) 111.

20. H. Ooguri, C. Vafa, 

`` Geometry of $N=2$ strings'' Nuc. Phys {\bf B 361} (1991) 469. 

21. G. Chapline, K. Yagamishi,  

`` Induced $4D$ self dual quantum gravity : a $W_\infty$ algebraic approach''. 

Class. Quantum. Grav. {\bf 8} (1991) 427. 

`` A Hamiltonian structure of the KP hierarchy , $W_{1+\infty}$ algebra and 

self dual gravity.''  Phys. Lett. {\bf B 259} (1991) 436.

22. Q.H. Park,  `` Extended conformal symmetry in real heavens `` 

Phys. Lett. {\bf B 236 } (1990) 429.

23. I. Bakas,  `` The large $N$ limit of extended conformal symmetries `` 

Phys. Lett. {\bf B 228 } (1989) 57.

24. D. Fairlie, J. Nuyts, `` Deformations and renormalizations of $W_\infty$ 

algebras `` Comm. Math. Phys. {\bf 134} (1990) 413. 

I. Bakas, E. Kiritsis, B. Khesin,  ``  The Logarithim Operator in Conformal 

Field Theory and $W_\infty$ algebras `` Comm. Math. Phys. {\bf 151} (1993) 233

25. A. Jevicki , `` Matrix Models, Open Strings and Quantization of 

Membranes `` hep-th/9607187. 

26. R. Zucchini, `` The Quaternionic Geometry of $4D$ Conformal Field Theory `` 

gr-qc/9707048.  

27. C. Castro, J. Plebanski : `` The Generalized Moyal Nahm  and 

Continous Moyal Toda Equations'' hep-th/9710041.

28. A. Strominger , `` Special Geometry `` Comm. Math. Phys {\bf 133} (1990) 163. 

29. E. Martinec, `` M Theory and N=2 Strings `` hep-th/9710122. 

30. C. Devchand, O. Lechtenfeld : `` On The Self Dual Geometry of N=2 Strings. 

hep-th/9801117.

31. S. Ketov, `` From N=2 Strings to F, M Theories `` hep-th/9606142. 

32. A. Lesov, G. Moore, N. Nekrasov and S. Shatashvili, : 

`` Four Dimensional Avatars  of two-dim RCFT `` hep-th/9509151. 

33. M. Bordemann , N. Neumeier, S. Waldmann : `` Homogeneous Fedosov Star 

Products on Cotangent Bundles I : `` Weyl Ordering with Differential Operators 

Representation `` q-alg/9707030. 

`` Homogeneous Fedosov Star Products on  Cotangent Bundles II : 

GNS Representations, the WKB expansion and Applications `` q-alg/9711016. 

P. Xu : `` Fedosov star products and quantum momentum maps `` q-alg/9608006. 

A. Astashkevisch, `` On Fedosov's quantization of semisimple coadjoint orbits'' 
MIT Ph.D Thesis. Math. Dept. 1996.  `

O. Kravchenko : `` Deformation Quantization of Symplectic Fibrations''.

math.QA/9802070.
                `` 

34. G. Dito, M. Flato, D. Sternheimer and L. Takhtajan : 

`` The deformation quantization and Nambu Mechanics ``  hep-th/9602016. 

35. M. Kontsevich, `` Deformation Quantization of Poisson Manifolds I `` 

q-alg/9709040.  

36. A. Voronov, `` Quantizing Poisson Manifolds `` q-alg/9701017. 

37. C. Castro, `` p-branes as Composite Antisymmetric Tensor Field Theories `` 

Int. Jour. Mod. Phys. {\bf A 13} (8) (1998) 1263-1292. 

38. I. Zois : `` On search for the M-theory Lagrangian `` hep-th/9703067. 

39. S. Lazzarini, `` Flat Complex Vector Bundles , the Beltrami Differential  and 

W- Algebras `` hep-th/9802083. 

40. M. Vasiliev : `` Deformed Oscillator algebras and higher spin gauge 

interactions of matter fields in  $2+1$ dimensions `` hep-th/9712246.

`` Higher-spin gauge theories in four, three, and two dimensions `` 

hep-th/9611024.

41. T. Ivanova , A. Popov, ``Some new integrable equations from the Self Dual 

Yang-Mills equations `` Phys. Lett {\bf A 205 } (1995) 158.                    `` 

A. Popov , ``Self Dual Yang-Mills symmetries and moduli spaces 

`` hep-th/9803183.   

T. Ivanova ,  `` On infinitesimal symmetries of the self dual Yang-Mills 

equations `` physics/9803028.

42. M. Saveliev, A. Vershik : `` Continual analogs of contragredient Lie 

Algebras `` Comm. Math. Phys. {\bf 126} (1989) 367. 

43. E. Nissimov, S. Pacheva, I. Vaysburd : `` Induced $W_\infty$ Gravity from a 

WZNW Model `` RI-92 ( Racach Institute) and BGU-92 ( Ben Gurion University) preprints.  

\bye

\centerline {Acknowledgements}
\smallskip

We thank  I.A.B Strachan for sending us the proof of how the  Moyal-Nahm equations admit reductions to the continuum Toda chain. 

\smallskip

25-O.F Dayi : ``q-deformed Star Products and Moyal Brackets `` q-alg/9609023. 

I.M Gelfand, D.B Fairlie : Comm. Math. Phys {\bf 136} (1991) 487. 

26-G. Rideau, P. Winternitz : Jour. Math. Phys. {\bf 34} (12) (1993) 3062.

27-J.R. Schmidt : Jour. Math. Phys. {\bf 37} (6) (1996) 3062.

28-V.G. Drinfeld : Sov. Math. Doke {\bf 32} (1985) 254.

29-M. Jimbo : Lett. Math. Phys. {\bf 10} (1985) 63.

30- G.W. Delius, A. Huffmann : `` On Quantum Lie Algebras and Quantum Root 

Systems ``q-alg/9506017.

G.W. Delius, A. Huffmann, M.D. Gould, Y.Z. Zhang  : `` Quantum Lie Algebras 

associated with $U_q(gl_n)$ and $U_q (sl_n)$'' q-alg/9508013.

V. Lyubashenko, A. Sudbery : `` Quantum Lie Algebras of the type $A_n$ `` 

q-alg/9510004.

31-M.Reuter : ``Non Commutative Geometry on Quantum Phase Space ``

hep-th/9510011. 

32-E.H. El Kinani, M. Zakkari : Phys. Lett. {\bf B 357} (1995) 105. 

J. Shiraishi, H. Kubo, H. Awata, S. Odake : `` A quantum deformation of the 

Virasoro algebra and Mc Donald symmetric functions `` q-alg/9507034 

33- C. Zha : Jour. Math. Phys. {\bf 35} (1) (1994) 517. 

E. Frenkel, N. Reshetikhin : `` Quantum Affine Algebras and Deformations of the 
Virasoro and $W$ algebras `` : q-alg/9505025. 

34- J. Mas, M. Seco : Jour. Math. Phys. {\bf 37} (12) (1996) 6510. 

35- L.C. Biedenharn, M.A. Lohe : `` Quantum Group Symmetry and $q$-Tensor

Algebras `` World Scientific, Singapore, 1995. Chapter 3. 

36- R. Floreanini, J. le Tourneaux, L. Vincent :  Jour. Math. Phys. {\bf 37}

(8) (1996) 4135.

37-G.Dito, M. Flato, D. Sternheimer, L. Takhtajan : ``Deformation quantization 

of Nambu Mechanics `` hep-th/9602016.

38- S. Albeverio,  S.M. Fei : `` Current algebraic structures over manifolds ,

Poisson algebras, $q$-deformation Quantization `` hep-th/9603114.

39- A. Connes : ``Non commutative Differential Geometry `` Publ. Math. IHES 

62 (1985) 41. 

40- B.V. Fedosov : J. Diff. Geometry {\bf 40} (1994) 213.  

H. Garcia Compean, J. Plebanski, M. Przanowski : ``Geometry associated with 

self-dual Yang-Mills and the chiral model approaches to self dual gravity ``

hep-th/ 9702046    

41- A. Jevicki : `` Matrix models, Open strings and Quantization of 

Membranes'' hep-th/9607187. 

T. Banks, W. Fischler, S.H. Shenker, L. Susskind `` M theory as 

a Matrix Model, a conjecture ``hep-th/9610043.

42- A. Marshakov : `` Nonperturbative Quantum Theories and Integrable Equations

`` ITEP-TH-47/96 preprint ( Lebedev Phys. Institute) 

H. Itoyama, A. Morozov : `` Integrability and Seiberg-Witten Theory'' 

ITEP-M 7- 95 / OU-HET-232 preprint. 

43-H.W. Braden, E. Corrigan, P.E. Dorey, R. Sasaki : Nuc. Phys {\bf B 338} 

(1990) 689.

L. Bonora, V. Bonservizi : Nucl. Phys. {\bf B 390} (1993) 205. 

P. Christie, G. Musardo : Nucl. Phys. {\bf B 330} (1990) 465. 

D. Olive, N. Turok, J. Underwood :  Nucl. Phys. {\bf B 409} (1993) 509.

T. Hollowood : Nucl. Phys. {\bf B 384} (1992) 523.

44- G.W. Delius, M.T Grisaru, D. Zanon :  Nuc. Phys {\bf B 382 } (1992 ) 365 .

45-P.E Dorey :  Nuc. Phys {\bf B 358 } (1991 ) 654 .

46- E. Frenkel : `` Deformations of the Kdv Hierarchy and related Soliton 

equations `` q-alg/9511003

47- H. Braden, A. Hone : `` Affine Toda solitons of the Calogero-Moser type ``

hep-th/ 9603178.

I. Krichever, A. Zabrodin : `` Spin generalization of the Ruijsenaars-Schneider 
model, nonabelian $2D$ Toda chain and representations of the Sklyanin algebra ``

hep-th/9505039.

\end